# The Formation of Disk Galaxies in a Cosmological Context:
## Structure and Kinematics

Matthias Steinmetz[1,2] (mhs@MPA-Garching.MPG.DE)

and

Ewald Müller[1] (ewald@MPA-Garching.MPG.DE)

[1] Max–Planck–Institut für Astrophysik, Karl–Schwarzschild–Straße 1, 85748 Garching b. München, FRG

[2]Institut für Theoretische Physik und Sternwarte der Universität Kiel, Olshausenstraße 40, 24098 Kiel, FRG

# Abstract

We present results concerning the internal structure and kinematics of disk galaxies formed in cosmologically motivated simulations. The calculations include dark matter, gas dynamics, radiative cooling, star formation, supernova feedback and metal enrichment. The initial model is a rigidly rotating overdense sphere with a mass of about $8 \, 10^{11} \, M_\odot$ which is perturbed by small scale fluctuations according to a biased CDM power spectrum. Converging, Jeans unstable and rapidly cooling regions are allowed to form stars. Via supernovae, metal enriched gas is returned to the interstellar medium.

From these initial conditions a galaxy forms which shows the main properties of spiral galaxies: a rotationally supported exponential disk which consists of young stars with about solar metallicity, a slowly rotating halo of old metal poor stars, a bulge of old metal rich stars and a slowly rotating extended halo of dark matter. Bulge, stellar and dark halo are supported by an anisotropic velocity dispersion and have a de Vaucouleurs surface density profile. The flattening of the dark and stellar halo is too large to be explained by rotation only. Whether the flattening of the bulge is caused by an anisotropic velocity dispersion or by its rotation cannot be answered, because of the limited numerical resolution due to gravitational softening. The velocity dispersion and the thickness of the stellar disk increase with the age of the stars. Considering only the young stellar component, the disk is cold ($\sigma = 20 \, \mathrm{km/sec}$) and thin ($z < 1 \, \mathrm{kpc}$). The dynamical formation process ends after about $4 \, \mathrm{Gyr}$, when the disk reaches a quasi–stationary state. During the subsequent $8 \, \mathrm{Gyr}$ mainly gas is transformed into stars decreasing the gas fraction from 40% to 10%. The star formation rate successively decreases during the quasi–stationary state from about $5 \, \mathrm{M_\odot/yr}$ at $z = 1$ to less than $0.5 \, \mathrm{M_\odot/yr}$ at the end of the simulation ($z = 0$).

## 1. Introduction

It is commonly believed that galaxies form by the collapse of gravitationally unstable, primordial density fluctuations. Within this general class of models, hierarchical cosmogonies seem to be favoured: First small structures form and subsequently larger structures develop by merging of smaller subunits (for a review see e.g., Silk & Wyse 1994). One example of such a cosmogony is the well known *Cold Dark Matter* (CDM) scenario. These kind of models can qualitatively reproduce many structural features observed in the cosmos. However, high resolution numerical simulations of the formation of galaxies, which allow one to compare these model predictions directly with observed galaxies, have still to be done.

Current models which try to explain the structure and kinematics of galaxies, particularly the dichotomy in the galactic morphology between spiral and elliptical galaxies, can roughly be divided into two classes: Models of the first class try to explain the different morphologies by means of a different star formation history, while models of the second class concentrate on differences in the initial conditions. Examples for the first class are models in which spirals are formed when the star formation is moderate and the gravitational collapse proceeds dissipationally. The formation of ellipticals involves an initial star burst which transforms most of the gas into stars on a short time scale the left–over gas being blown out by energy input due to supernovae. After the starburst the remaining stellar system collapses dissipationlessly. Simulations of van Albada (1982) seem to support that due to violent relaxation such ellipsoidal structures indeed form, as long as the initial conditions are not too regular or symmetric. A typical representative of the second class of models is the merging scenario, in which at least some of the elliptical galaxies form by the coalescence of two colliding galaxies.

The two classes of models also reflect the two main problems of galaxy formation, namely our poor understanding of star formation and the arbitrariness of the initial conditions. One has tried to overcome the latter problem by extending cosmologically large scale N–body simulations to smaller scales including the effects of gas dynamics and star formation (Navarro & Benz 1991, Katz & Gunn 1991, Katz 1992, and Navarro & White 1993, 1994).

In this publication we describe numerical simulations starting from the same initial conditions as in Katz (1991). The simulations include the dynamical and chemical evolution of gas, dark matter and stars including radiative cooling. In a recent publication (Steinmetz & Müller 1994a, hereafter SM2) we have shown, that in such simulations a disk galaxy forms, which shows the main chemical properties of spiral galaxies: a metal poor halo composed of old stars with nearly no metallicity gradient, a metal rich bulge of old stars and a disk component of young stars of about solar metallicity. The disk stars show a metallicity gradient of typically $d \log Z/dr = -0.05 \, \text{kpc}^{-1}$. In the following we present a detailed investigation of the structure (section 3) and the kinematics (section 4) of the different components. We shall demonstrate, that stars identified chemically or by age to be Pop I or Pop II stars, also show the observed distribution and the dynamical properties of the respective population.

## 2. Methods and global evolution

### 2.1 Initial conditions

The most realistic way to set up initial conditions for simulations of the formation of galaxies is that of Navarro & Benz (1991) and Navarro & White (1994), which follow the evolution of regions with a comoving radius of a few Mpc with proper inclusion of the tidal effects and the mass exchange with the surrounding material. However, since the total mass of such a region can exceed $10^{13} \, \text{M}_\odot$, and because the total particle number is limited to several ten thousands on current supercomputers, the numerical resolution per forming galaxy of mass $\lesssim 10^{12} \, \text{M}_\odot$ is limited to several hundreds or at most a few thousands of particles.

Katz & Gunn (1991) consider an isolated sphere on which small scale fluctuations according to a CDM power spectrum are superposed. To simulate the influence of fluctuations of larger wavelengths, the sphere is initially set into rigid rotation with a spin parameter $\lambda \approx 0.08$. In addition, the density of the sphere is slightly enhanced above the background density to guarantee the collapse of the sphere. Starting from these initial conditions, a resolution up to a few $10^4$ particles per galaxy can be achieved, but mass exchange across the boundaries, which is of particular importance at lower redshifts, is completely neglected.

At first glance, the initial conditions of Katz & Gunn (1991) seem to be quite artificial. However, they can be physically motivated. Within the framework of first order Lagrangian perturbation theory (Buchert 1992), the particle trajectories at early epochs are essentially described by the tidal tensor $\partial_i \partial_j \Phi$, where $\Phi$ is a quantity proportional to the gravitational potential. The tidal tensor can be decomposed into a diagonal and a traceless part (Hoffman 1986). Its diagonal part represents Poisson's equation and is responsible for a contraction (or expansion for negative density contrasts) of the considered system. This effect is approximated by enhancing the density of the sphere. The traceless part, which corresponds to a shear (not a rotation!), induces angular momentum (White 1984, Steinmetz & Bartelmann 1994). This part is simulated by setting the system into rigid rotation. Obviously, this ansatz bears three inconsistencies: (i) The linear theory predicts that angular momentum is acquired by a shear field (i.e., a symmetric tensor) without vorticity, which cannot be modelled by setting the system into rotation (i.e., an antisymmetric tensor). (ii) Linear theory predicts, that a spherically symmetric volume does not acquire any angular momentum at all. (iii) We consider the dynamical evolution far into the non linear regime, i.e., crossing of particle trajectories (which correspond to mass transfer or merging) becomes important even on the mass scale of a galaxy.

Although the quasi–cosmological initial conditions of Katz & Gunn (1991) have a physical basis (especially for the early evolution), they are quite special. They may be typical for field galaxies or for galaxies which forms in an open universe ($\Omega < 1$), where further clustering ceases at $z \approx 1/\Omega - 1$. In spite of these deficiencies the general outcome of simulations using the Katz & Gunn (1991) initial conditions seems to agree quite well with that of the simulations performed with realistic boundary conditions. Furthermore, the dark haloes formed in dark matter simulations of Katz (1991), who used the simplified initial conditions, are in good agreement with large scale simulations performed by Frenk et al. (1987).

The numerical method on which our models rely is a combined N–body–SPH treecode. Every particle is propagated on its own time step by means of a hierarchy of time bins. The code is described and tested in detail in Steinmetz & Müller (1993, 1994b, henceforth SM1 and SM3). Recently, the code has been modified to a direct summation N–body-SPH code, where the gravitational interaction is solved on the special purpose hardware GRAPE (for a review see Sugimoto 1993). We start at a redshift $z = 25$ with a homogeneous, rigidly rotating sphere with a mass of $8 \, 10^{11} \, M_\odot$, a radius of $50 \, \mathrm{kpc}$ ($\hat{=} 0.65/h \, \mathrm{Mpc}$ comoving) and a spin parameter of $\lambda \approx 0.08$, the latter being typical for dark haloes formed in hierarchical cosmogonies (Barnes & Efstathiou 1987, Steinmetz & Bartelmann 1994). Small scale fluctuations are imposed by displacing the particles from their regular grid positions and assigning them velocities according to the Zel'dovich approximation (Zel'dovich 1970). The combination of mass and radius roughly corresponds to a $\delta_\mathrm{i} = 0.27$ overdensity relative to the cosmic background density of a $\Omega = 1, H_0 = 50 \, \mathrm{km/sec/Mpc}$ universe. Therefore, it can be identified with a $3\sigma/b$ overdensity of a biased CDM spectrum on a mass scale of $8 \, 10^{11} \, M_\odot$, where $b$ is the biasing parameter. In contrast to Katz (1991) the Hubble expansion of the sphere is reduced by a factor $(1-\frac{1}{3}\delta_\mathrm{i})$ to match the $\delta = -\nabla \mathbf{v}$ condition from linear theory (Peebles 1980, Bartelmann et al. 1994). The sphere reaches the turnaround point (i.e., its maximum expansion) at a redshift of $z \approx 5.5$.

Initially, we assume baryons and dark matter to be equally distributed. The baryons are in form of gas which consists of 76% H and of 24% He by mass. The gas is cold , i.e., it has the temperature of the radiation background (70K), and no stars have formed yet. The further evolution of the baryonic component differs from that of the dark component, because it suffers adiabatic compression, shock heating and radiative cooling. Gas regions which are convergent, Jeans unstable and rapidly cooling can form a star particle in a way similar to that described in Katz (1992) or in Navarro & White (1993). The mass and phase space coordinates of the newly formed star particles are given by a weighted sum over the neighbouring gas particles according to the SPH algorithm. Afterwards the star particles are treated as additional collisionless particles. We assume that the stars, which are represented by a star particle of $\approx 10^7 \, M_\odot$, are distributed according to a Miller–Scalo (1979) initial mass function with a time independent upper and lower mass limit of $100\,\mathrm{M}_\odot$ and $0.1\,\mathrm{M}_\odot$, respectively. Stars more massive than $8\,\mathrm{M}_\odot$ become type II supernovae, each of which deposits $10^{51}$ erg in form of thermal energy in the gas. Momentum input by supernovae as in Navarro & White (1993) is not yet included. Except for a $1.4\,\mathrm{M}_\odot$ remnant all mass of a massive star is returned metal enriched to the gas component. For details of the star formation algorithm we refer to SM2 and SM3. The gravitational softening is taken to be 3 kpc for the dark matter, 1.5 kpc for the gas and 1 kpc for the stellar component, respectively. Thus it scales with the particle mass $m$ roughly like $m^{1/3}$. Numerical studies and test simulations in SM1 and SM3 imply, that the results of the simulations are converged within the resolution limit of about 1–2 kpc.

## 2.2 Data reduction

As in SM2, we identify stellar populations mainly by their age: Young stars are called those which form during the last 0.5 Gyr, old stars those which form during the first 2 Gyr of the simulation, i.e., which are older than 11 Gyr. For reasons which will become clear in the next sections, stars younger than 8.5 Gyr are called disk stars. Moreover, we will use the synonyms bulge and halo for old stars with a metallicity larger and smaller than the solar metallicity, respectively. According to our previous results (SM2), the bulge stars are concentrated near the center of the galaxy, whereas the halo stars are preferentially found at large distances to the galactic plane.

To analyze the structure and kinematics of a final configuration, we have binned it into cylindrical shells coaxial to the rotation axis. The thickness of the 20 – 50 shells is selected in such a way that (i) the number of particles per shell is roughly constant, and (ii) the thickness of the shells varies smoothly with distance. The shell thickness ranges from about 0.2 kpc near the center to 5 kpc in the outskirts of the galaxy. On average, about 80 gas, 80 dark matter and 400 star particles are contained in each shell. Subsamples are binned in such a way, that, if possible, at least 50 stars contribute to each bin. The resolution, therefore, is sufficient to obtain information on the velocity dispersion with an accuracy better than 10%. Discriminating populations the accuracy is better than 30%.

To investigate the reliability of local properties derived for the final configuration of a simulated galaxy, we use a statistical bootstrap as recently proposed by Heyl et al. (1994). Their main idea can be outlined as follows: Suppose one has an ensemble of N particles resulting from a N–body or a SPH simulation. From that ensemble one creates a series of new ones by drawing $N$ integer random numbers equally distributed in the interval $[1, N]$. These $N$ random numbers provide the indices of all those particles belonging to the new ensemble, i.e., one may find some particles several times in the new ensemble, and some other particles not at all. One can show, that in the statistical mean $1/e \approx 37\%$ of the particles of the original ensemble do not belong to a bootstrapped sample. A feature, which only occurs in a few ensembles is, therefore, probably artificially produced by only a few particles, whereas a feature which occurs in every ensemble is statistically much more significant. By comparing different ensembles one can get an error estimate for properties derived from the particle distribution, as e.g., the scale length of the disk or the effective radius of the bulge–halo system etc. For more details about statistical bootstrap we refer to Heyl et al. (1994) and to Efron (1982) and Efron & Tibshirani (1986).

To get rough information on the luminosity evolution of a galaxy and on the mass to light ratio, we weight a star particle with its (time dependent) mean luminosity, which can be derived from the IMF assuming main sequence luminosities according to a mass luminosity relation (Kippenhahn & Weigert 1991). For a more detailed description we refer to SM3.

## 2.3 The simulations

The simulations start with 4000 gas and dark matter particles. At the end of a simulation, typically 30 000 star particles have formed. In contrast to SM2, where the simulations were stopped at $z = 1$, we are now able to continue the simulations up to $z = 0$ using GRAPE. A simulation requires about 1300 system time steps of $10^7$ yr and 60 000 time steps on the smallest time bin of the multiple time step scheme (see above and SM1). Roughly 2/3 of the time steps are necessary to compute the quasi–stationary evolution from $z = 1$ to $z = 0$. The simulations have been performed on a SPARC 10 workstation connected to the N–body hardware integrator GRAPE (Sugimoto 1993, Steinmetz 1994). One simulation requires about 150 h of CPU time, 110 h of which are necessary to evolve the quasi-stationary state from $z = 1$ to $z = 0$. The simulations are bottlenecked by the limited speed of the workstation, which carries 80% of the computational load. A similar simulation on a CRAY YMP with a SPH tree code requires 60 h CPU time to proceed up to $z = 1$. The highly clustered state at the end of the simulation (half of the 30000 stars can be found in a 5 kpc sphere) implies a large interaction list for the tree code. We estimate that a simulation up to $z = 0$ would require more than 500 h on a Cray YMP.

The correlations between age, metallicity and distribution of the stars found in SM2 in simulations up to $z = 1$, are not altered when extending the simulations to $z = 0$. Metal poor old stars ($t_* > 10.5$ Gyr) are found in an extended halo, metal rich old stars are located close to the center of the galaxy, and the disk mainly consists of stars with $t_* < 8.5$ Gyr possessing about solar metallicity. The dynamical evolution of a galaxy which is mirrored in the star formation history (Fig. 2), can be summarized as follows (for a detailed description we refer to Katz (1992) and SM3): The largest density maxima in the initial model becomes nonlinear at a redshift of $z \approx 10$. They collapse and the first stars are formed. At $z = 5.5$ the turnaround point of the sphere is reached, i.e., it begins to collapse. At $z \approx 3$ several still gas rich density maxima merge resulting in the first star burst with a star formation rate of $30 \, M_\odot/\text{yr}$ at $t \approx 11.5$ Gyr (see Fig. 2). Subsequently, the star formation rate drops by a factor of two. A second star burst occurs at $t \approx 10$ Gyr when the disk component forms. The star formation rate in this second burst again reaches a value of $30 \, M_\odot/\text{yr}$. Whereas the first star burst is mainly powered by the violent merging of a relatively small fraction of particles, the second star burst results from a moderate but coherent star formation activity in a large gas fraction induced by the overall collapse of the sphere.

At $z \approx 1$ the formation of the disk is almost finished and only 40% of the galaxy mass is still in form of gas. The star formation rate at this epoch is about $5 \, M_\odot/\text{yr}$. Later in the evolution the dynamical properties and the density distribution remain nearly unchanged, only gas is continuously transformed into stars. Due to the exhaustion of the gas the star formation rate is steadily decreasing, and the small star formation and supernova rates do no longer affect the galactic structure. Obviously, this statement only holds for the kinematics and the mass distribution of the galaxy. The luminosity of the galaxy can still change significantly due to the death of luminous high mass stars. At $z = 0$, the gas fraction of the disk has dropped to 15% and the star formation rate has reached a value of about $0.5 \, M_\odot/\text{yr}$. The late evolution ($t < 10$ Gyr), therefore, is qualitatively similar to that of the model proposed by Wang and Silk (1994), where the galactic evolution is approximated by a stationary exponential disk in which Toomre unstable regions form stars on a gravitational time scale.

| component | $z_{25}$ [kpc] | $z_{50}$ [kpc] | $z_{95}$ [kpc] | $\sigma_z$ [km/sec] |
|---|---|---|---|---|
| gas | 0.06 | 0.4 | 3.0 | - |
| stars | 0.25 | 0.8 | 5.0 | 70 |
| stars $t < 2$ Gyr | 0.4 | 0.9 | 3.0 | 20 |
| stars $t > 10$ Gyr | 0.3 | 0.8 | 6.0 | 100 |

Table 1:
Thickness of the disk $z$ and vertical velocity dispersion $\sigma_z$ for different components. Shown are the 25 (50, 95) percentile values, i.e., 25%, (50%, 95%) of the mass of a component can be found within a distance of $z_{25}$ ($z_{50}$, $z_{95}$) perpendicular to the disk.

The distributions of gas and stars of the final configuration show (see Fig. 3 and Fig. 4) that the remaining gas settles into a thin disk and that the stars also form a disk–like object possessing however a much larger thickness. The disk of all those stars which form in the quasi–stationary state ($t < 8.5$ Gyr) is remarkably thinner, although still much thicker than the gas disk. The distribution of the youngest stars ($t_* < 1$ Gyr) is similar to that of the gas except for being slightly thicker. The old stars, which form during the first star burst do not form a disk but instead an ellipsoidal system. As already pointed out in SM2, discriminating the old stars by their metallicity shows that old metal poor stars form a diffuse halo (two middle panels in Fig. 4), while old metal rich stars are concentrated strongly near the very center of the galaxy forming an ellipsoidal bulge (lower two panels in Fig. 4).

## 3. Structure of disk galaxies

If the history of elliptical galaxies is characterized by hierarchical merging involving lumps which mainly consist of stars, then the formation of dark haloes and elliptical galaxies is similar. Hence, it is valuable to compare the structures of both objects (Zurek et al. 1988), although ellipticals are much more compact than dark haloes. The surface luminosity of ellipticals is well described by a de Vaucouleurs (1948) law ($\log I \propto r^{1/4}$), which as recently shown by Burkert (1993) is also fulfilled for radii less than the effective radius $r_e$. However, most models, which describe the formation of ellipticals via dissipationless collapse or via merging spirals fail to obey the $r^{1/4}$–law for $r < r_e$ (Burkert, in preparation). Starting from identical initial conditions the surface density of the dark halo obtained in both a simple N–body simulation and in a simulation including gas dynamics and star formation is shown in Fig. 5. It is also well described by a $r^{1/4}$–law, the effective radius being $r_e = 30$ kpc (35 kpc) for the simulation with (without) gas (Fig. 5). Note, that the surface density profile also follows the de Vaucouleurs law for radii less than the effective radius for both the model with and without gas. Furthermore, the surface density of the halo is remarkable higher in the simulation with gas and the corresponding radial profile agrees much better with a $r^{1/4}$–law at smaller radii.

Katz (1992) noted, that in simulations with and without star formation the gas surface density fulfills an exponential law, whereas no such clear trend is visible for the stellar components. He also observed that the scale length of the stars is smaller than that of the gas, which is confirmed by our simulations (see first panel of Fig. 6). Obviously, the total mass distribution of the stars is only poorly fitted by an exponential, especially for the two early models (at $z = 1.45$ and $z = 0.78$). The deviation is large in the innermost layers ($r < 3\,\text{kpc}$) but acceptable in the outer layers ($r > 10\,\text{kpc}$), where we find a scale length of 4.5 kpc. At late times ($z = 0$) the surface density of the stars is well represented by an exponential disk with a scale length of 5.5 kpc. The combined surface density of the gas and the stars shows much less evolution between $z = 1.48$ and $z = 0$ compared to the stellar disk (upper right panel of Fig. 6).

This behaviour becomes even more evident, if we consider the system made of gas and disk stars, only. Except for the very center, the surface density of the gas component at $z = 1.45$ is identical to the surface density of the disk stars at $z = 0$, both showing a clear exponential profile with a scale length of 6.5 kpc (upper and lower left panel of Fig. 6). It is also identical to the surface density of the sum of gas and disk stars, which remains practically unchanged for $1.45 \geq z \geq 0$. This fact strongly supports the interpretation, that the system has reached a stationary state, where only gas is continuously transformed into stars. The evolution of the gas surface density furthermore shows, that the gas is depleted in the inner regions in a way similar as described analytically by Wang & Silk (1994). Their model is based on an axisymmetrical exponential gas disk in which stars are formed on a gravitational time scale in Toomre unstable regions.

The surface density of the old stellar component ($t > 11\,\text{Gyr}$) representing the halo and bulge follows a de Vaucouleurs law with an effective radius of 2–3 kpc, which does practically not change during the late evolution (lower right panel of Fig. 6). Note that the $r^{1/4}$–law is also fulfilled inside the effective radius of about 2 kpc. We have to point out, however, that the results for the bulge component may be influenced by the limited numerical resolution due to gravitational softening (1 kpc). Concerning the compactness of the bulge we refer to the discussion in section 5.

Since the old and young stellar component are well described by a $r^{1/4}$- and an exponential law, respectively, we have tried to fit the surface density with a disk+bulge model. Indeed, a combination of a disk with a scale length of $r_\text{d} = 5.8\,\text{kpc}$, a bulge with $r_\text{e} = 1.7\,\text{kpc}$ and a disk to bulge ratio of 1.3 to 1 seems to provide a reasonable fit. Weighting the stellar mass with a luminosity function (see above) we have determined the surface brightness of our models at $z = 0$, which is well described by a disk–bulge combination with $r_\text{d} = 5.7\,\text{kpc}$ and $r_\text{e} = 2.3\,\text{kpc}$ and a disk to bulge ratio of 3:1, i.e., the modelled galaxy has the typical properties of a spiral galaxy of Hubble class Sb. By means of the bootstrap analysis we find for the surface density a mean (maximum) error of 2% (6%) for $r_\text{d}$, 15% (40%) for $r_\text{e}$ and 15% (40%) for the disk to bulge ratio. The errors for the surface brightness are a factor of 1.5 larger.

Concerning the thickness of the stellar and gas disk one sees from Tab. 1 that the collapse of the gas perpendicular to the disk cannot be stopped until nearly all gas is accumulated in a disk with a thickness of about one softening length. Roughly half of the gas mass is contained in a thin disk with a thickness of about one third of a softening length, while 95% (50%) of the stellar mass can be found within a disk with a thickness of about 5 kpc (0.8 kpc). Considering only young stars, the 50 percentile value remains almost unchanged, whereas the 95 percentile value decreases to 3 kpc (see Tab. 1). The rather invariant behaviour of the 25 and 50 percentile value is artificial, because the height of all these stars is less than the gravitational softening length. Fig. 7 shows that the mean latitude is roughly 1 kpc for young stars and increases for older stars reaching $4-5$ kpc for stars within an age of 10 Gyr, which are probably the oldest disk stars. The large latitude of the very old stars ($t_* > 10$ Gyr) indicates, that they form the stellar halo. Since the gas disk is thinner than the stellar one, we argue, that in a simulation with a much higher spatial resolution, e.g., with a gravitational softening of 0.1 kpc, the mean latitude of the gas and of the youngest stars would be smaller than 1 kpc.

We conclude the discussion of the disk and bulge structure with an investigation of the mass to light ratio $\Upsilon$, keeping in mind that the luminosity of a star is assumed to be its main sequence value. Considering only the baryonic mass (stars and gas) the mass to light ratio $\Upsilon = 5 \ldots 6\,\Upsilon_\odot$ at $z = 0$. Since the scale length of the surface luminosity is larger than that of the surface density, $\Upsilon$ decreases with increasing radius. At 8.5 kpc, which corresponds to the distance of the sun from the galactic center, $\Upsilon \approx 5\,\Upsilon_\odot$, while at a distance of 20 kpc, $\Upsilon \approx 2\,\Upsilon_\odot$. Beyond 20 kpc, the mass to light ratio does not further decrease. At $z = 1$, when the fraction of luminous, massive young stars with $\Upsilon_{\text{star}} \ll 1$ is much larger than at $z = 0$, the mass to light ratio is lower, namely $3\,\Upsilon_\odot$ in the center and $1\,\Upsilon_\odot$ at 20 kpc. To consider the effect of the dark halo mass on the mass to light ratio, we took into account all mass of the dark halo within a slice of thickness 1 kpc complanar to the galactic disk. Then we obtain a mass to light ratio of $15\,\Upsilon_\odot$ in the center. The ratio decreases with radius to a value of about $7\,\Upsilon_\odot$ at a radius of 15 kpc. Further out the mass to light ratio remains roughly constant up to distances of 25 kpc. Beyond that radius $\Upsilon$ increases to a value of more than $30\,\Upsilon_\odot$ at a radius of 40 kpc. All these values seem to be at least in qualitative agreement with observations of our galaxy (see e.g., Binney & Tremaine 1987). Note, that the inclusion of dark matter does not change the mass to light ratio in the inner 25 kpc by more than 50%. For an observer inside such a model galaxy it would, therefore, not be possible to infer only from the internal galactic dynamics the existence of dark matter in an amount that it is consistent with the nucleosynthesis constraint. The situation for such an fictitious observer is similar to that of an observer in our milky way today.

## 4. Dynamics of disk galaxies

In the last section we have shown that there exists a correlation between the age and metallicity of stars and their spatial distribution, thus defining different stellar populations. In this section we will investigate to what extent these different populations can also be identified by their dynamical properties. Fig. 8 shows the rotation velocity for the different populations as a function of radius together with the circular velocity $v_{\text{c}}$ (see Binney & Tremaine 1987) defined by

$$v_{\text{c}}^2 = r\frac{\partial \Phi}{\partial r} \approx \frac{G\,M}{r}, \qquad (1)$$

where $r$ is the distance from the rotational plane. While the approximation in Eq. (1) is quite accurate in case of a spherically symmetric mass distribution, the circular velocity can be up to 15% larger than that estimated from a spherical mass distribution with the same radial mass distribution $M(r)$ for an exponential disk. A rotationally supported configuration should, therefore, have a rotation velocity close to the circular velocity. To discriminate the effects due to the potential of baryonic and dark matter, we also show the circular velocity resulting from the baryons, only. The velocity dispersions $\sigma_r$, $\sigma_\phi$ and $\sigma_z$ are shown in Fig. 1. together with the circular velocity of the total mass distribution. From the upper panel of Fig. 8 it is obvious that the gas component practically rotates with the circular velocity, i.e., it is rotationally supported. The rotation velocity steeply rises to a value of about 250 km/sec in the innermost 4 kpc and then remains practically constant up to distances of 30 kpc. This value is a factor of 1.7 larger than the circular velocity derived from the baryonic mass distribution. The rise in the inner 4 kpc is probably affected by gravitational softening, i.e., the rotation velocity would probably rise even more steeply in a better resolved simulation.

The rotation velocity of the dark component $\bar{v} < 0.25 v_c$, i.e., the dark halo has to be stabilized by velocity dispersion. Indeed, we find $\sigma \approx 150$ km/sec, which is sufficient to stabilize the dark halo. Concerning the orientation and the shape of the dark halo our results show that it is only slightly disaligned with respect to the disk and possesses an ellipsoidal shape. The three principle axes are $a_1 = 30$ kpc, $a_2 = 31$ kpc and $a_3 = 21$ kpc, where

$$a_j = \sqrt{\frac{1}{5}\mu_j}, j = 1, 2, 3, . \qquad (2)$$

The quantities $\mu_j$ are the three eigenvalues of the moment of inertia tensor of the dark halo. Hence, the ellipticity of the dark halo $\epsilon = 1 - a_3/a_1 \approx 1 - a_3/a_2 = 0.3$. Because $a_1$ and $a_2$ are perpendicular to the rotation axis of the disk, the dark halo is oblate. To explain this degree of flattening by rotation, $\bar{v}/\sigma \gtrsim 0.6$, whereas a comparison of Figs. 8 and 1 suggests $\bar{v}/\sigma \lesssim 0.4$, i.e., the halo can only be partially flattened due to rotation. On the other hand, the velocity dispersion of the halo is anisotropic with $\sigma_r/\sigma_z \approx 1.2$. According to Binney & Tremaine (1987, Figs. 4-5 and pages 216–217), this combination of anisotropy and amount of rotation $\bar{v}/\sigma$ is consistent with a flattening $\epsilon = 0.3$.

Since the star formation rate is only of the order of 0.5 $M_\odot$/yr during the last Gyr, only a few hundred particles are formed during that period. This obviously is too small a number to determine the velocity dispersion. Therefore, we make use of the quasi–stationarity and analyze a model at $z = 0.8$. It is then possible to analyze the class of newly formed stars which are formed during the last 0.5 Gyr prior to $z = 0.8$. The star formation rate at $z = 0.8$ is equal to $\approx 4\,M_\odot$/yr, i.e., a few thousand stars are formed during that epoch.

The rotation velocity is similar to that of the gas, but the velocity dispersion is small $\sigma \approx 20$ km/sec (see Figs. 8 and 1 the disk is cold. For subsamples of stars of age $[0, 3]$ Gyr and $[3, 8.5]$ Gyr, we find a mean velocity dispersion of 40 and 80 km/sec at a distance of 10 kpc. In all cases, $\sigma_r$ is about 1.5 times larger than both $\sigma_z$ and $\sigma_\phi$. For comparison we mention that Fall (1981) and Edvardsson et al. (1994) obtained $\sigma = 10$ km/sec for the youngest stars ($< 1$ Gyr), $\sigma \approx 20$ km/sec for stars of age $1 < t < 3$ Gyr and $\sigma_r = 50$, $\sigma_\phi = 30$, and $\sigma_z = 25$ km/sec for the oldest M and K stars. Although we can confirm the general trend that the disk becomes warmer with increasing age of the stars, the velocity dispersions tend to be too large in our model. We point out, that newly formed stars have nearly no velocity dispersion. But due to interaction with massive objects the disk is probably heated up (Wielen 1977), thus continously increasing the velocity dispersion of the stars. Because of the poor mass resolution (compared to stellar masses) and the small particle number (compared with the number of stars in a galaxy), this heating mechanism is probably overestimated in the numerical simulation and may, therefore, cause the too rapid increase of the stellar velocity dispersion in our models.

The dynamical properties of old stars (i.e., the bulge halo system) are are similar to that of the dark halo. The old stars are only slowly rotating (50 km/sec) and their velocity dispersion is large (100 –150 km/sec) and anisotropic (see Figs. 8 and 1). These results are consistent with observations of the halo and bulge of our galaxy (Fall 1981, Zinn 1985, de Zeeuw 1992), although the bulge of our galaxy seems to rotate faster (100 km/sec; de Zeeuw 1992). Both the anisotropy and the flattening of the bulge halo system is larger than that of the dark halo, the ratio of the principle axis being $a_2/a_1 = 0.85$ and $a_3/a_1 = 0.65$, respectively.

The analysis of the numerical results is complicated by two problems. Firstly, it is difficult to discriminate between bulge stars and the first generation of disk stars which are formed near the center (a problem also encountered by an observer). Secondly, the rotation velocities in the bulge, i.e., in the innermost 2 kpc of the galaxy, are affected by gravitational softening. We find that whereas the dynamical properties of the halo component are relatively insensitive against changes in the age criterion (20% error), those of the bulge can change by more than 50%. We, therefore, cannot answer the question, whether the bulge is rotationally flattened or not.

## 5. Summary and Discussion

We have presented three dimensional numerical simulations of the formation of disk galaxies including gas, dark matter, star formation and a radiative cooling. Although the initial conditions are not yet fully consistent with the predictions of hierarchical structure formation, they are suitable to disk galaxies formed in the field or in an open universe. Moreover. our results demonstrate that it is possible to simulate the formation of disk galaxies with a spatial resolution allowing for statements about local properties down to scales of kpc size.

The main result of our simulations is that we are able to identify the major components of spiral galaxies, namely disk, bulge, stellar halo and dark halo using a combined age–metallicity criterion. The disk can be discriminated from the bulge-halo system by means of a simple age criterion, because the bulge-halo system is formed during the first starburst triggered by the collapse of small scale density fluctuations, whereas the disk system forms later in the evolution during the collapse (and infall) of fluctuations on larger scales. As already shown in a previous publication (SM2), the first generation of stars can be further divided into old metal poor and old metal rich ones forming the halo and the bulge, respectively. The metal rich bulge stars preferentially form in the region, where the small scale fluctuations have maximum amplitude, whereas the metal poor halo stars stem from a much more extended region covering to the secondary maxima of the small scale fluctuations. The density distribution as well as the chemical and dynamical properties of the individual components agree very well with what is known from observations. Using a bootstrap analysis and varying the classification criterion for the different stellar populations we find, that the properties derived for the disk are accurate up to about 10%, whereas the properties of the stellar bulge and halo are much more effected by the limited spatial resolution and by the limited particle number. However, in most cases the accuracy is better than 30-50%.

The dark halo is slowly rotating, supported by velocity dispersion, has an oblate shape, and is aligned with the rotational plane of the initial model. The flattening of the dark halo ($\epsilon = 0.3$) is too large to be solely explained by rotation. However, it is consistent when also taking into account the anisotropy of the velocity dispersion of our models. Over a large radial intervall ranging from 2 kpc up to 100 kpc the surface density of the dark halo is well described by a de Vaucouleurs profile with an effective radius of about 30 kpc. In comparison with simulations neglecting gasdynamical effects, the halo is more centrally condensed.

The stellar halo has a size of about 30 kpc, and is supported by its velocity dispersion of about 100 km/sec. Its density profile can also well be described by a de Vaucouleurs profile. The stellar halo is triaxial with an axis ratio of 1/0.85 and 1/0.65, respectively. It is also aligned to the rotational plane its minor axis being roughly parallel to the rotational axis of the disk. The stellar halo rotates slowly, i.e., its flattening is caused by the anisotropic velocity dispersion.

The surface density of the bulge can also be described by a de Vaucouleurs profile with an effective radius of 2–3 kpc the $r^{1/4}$–law holding inside the effective radius, too. Its rotation velocity is smaller than 100 km/sec and its velocity dispersion is $\approx$ 100 km/sec.

At first glance the compactness of the bulge comes as a surprise, because the collapse of a dissipationless system at redshift of about $z = 3\ldots 4$ should lead to a much more extended system. The compactness of the bulge, therefore, implies that dissipation also played an important role during the formation of the bulge component. The assumption that the collapse proceeds dissipationally is not unreasonable due to the construction of the star formation law: Only 33% of the mass of a gas particle can be transformed into stars in a star–formation time step of 10 million years. Consequentely, it needs several tens of million years to transform all the gas into stars (for a detailed description and discussion of the star formation algorithm see also Katz 1992 and SM3), i.e., the gas consumption time is comparable to if not larger than the collapse time for a mass scale of about $10^{10}$ M$_\odot$. Hence, the gas can radiate off a remarkable amount of energy which results in the compact bulge found in our simulations.

However, since the effective radius of the bulge is only 2–3 times the gravitational softening and since it consists only of about 2000 particles, one must carefully check for possible numerical artifacts which may falsify the results on these scales. Typically three different resolution related numerical problems can be encountered: (i) The direct influence of the gravitational softening, which implies that the maximum phase space density a system can adopt is limited. (ii) Two–body relaxation effects may not be sufficiently suppressed by gravitational softening, i.e., on timescales

$$t_{\rm relax} = \frac{0.1\,N}{\ln N} t_{\rm cross}, \qquad (3)$$

where $N$ is the particle number and $t_{\rm cross}$ the local crossing timescale, the system will contract in the central region (Binney & Tremaine 1987). (iii) The standard Monaghan & Gingold (1993) artificial viscosity has a non vanishing shear component. The shear viscosity can transport angular momentum leading to a steady mass flow towards the bulge. The first of these problems is not critical for the bulge compactness, because if this effect is important, it would predict an even more compact bulge in a better resolved simulation. Concerning the two–body relaxation effects, one has to be more carefully. For the typical velocities in the bulge (100 km/sec) and its typical extension (2 kpc), the local dynamical time scale is 20 million years, i.e., for the bulge ($\approx$ 2000 particles) on gets a critical time scale of $t_{\rm relax} \approx 0.5$ Gyr. This is more than an order of magnitude smaller than the Hubble time. However, as shown in section 3, the surface density of the disk+bulge system is remarkably constant within the last 9–10 Gyr. Therefore, relaxation effects are sufficiently suppressed by gravitational softening. Finally, concerning the non–vanishing shear compo-nent of the Monaghan & Gingold (1993) artificial viscosity, we have increased the viscosity coefficients by a factor of two in one simulation. In another one, we have used the artificial viscosity proposed by Balsara (1994). It is found that the latter modification reduces the mass flow toward the galactic center by several orders of magnitude (Englmaier, Gerhard, & Steinmetz, in preparation, see also SM3). Using both modifications we obtain a quantitatively similar bulge component, which suggests that effects due to the artificial viscosity are indeed small.

To strengthen this conclusion, we have also performed other simulations with different amounts of small scale power. It turns out, that models without any small scale power do not show a bulge at all, whereas models with more small scale power (i.e., with an initially larger amount of kinetic energy in random motions) show a more massive bulge. The code, therefore, is principally able to simulate much less pronounced bulge components, too. Hence, we have good reasons to conclude that the bulge is not a numerical artifact and to believe that its properties are at least qualitatively well described.

The disk has an exponential profile and is rotationally supported. The scale length of disk (stars and gas) is $\approx 6.5$ kpc and does not significantly change during the last 10 Gyrs. The dynamical evolution of the disk is almost finished 3 Gyr after the turnaround time (i.e., the time of maximum expansion) for a galactic scale. During the last 8–9 Gyr the disk is in a stationary state, where only gas is continuously transformed into stars. The gas fraction declines from 40% to less than 15%, while the star formation drops from $5\,\mathrm{M_\odot/Gyr}$ to less than $0.5\,\mathrm{M_\odot/Gyr}$. The quiescence during the last 8 Gyr is, however, partially an artifact of the assumed vacuum boundary condition. The thickness of the disk is essentially limited by the numerical resolution. Nevertheless, the general trend is correctly described, namely that stars form in a thin disk ($< 1\,\mathrm{kpc}$) and become heated up with time. Consequently the thickness of the disk as well as the vertical velocity dispersion increase with the age of the stars. The heating mechanism in our model galaxy seems to be more efficient than in a typical spiral galaxy. We argue that this is an artifact caused by the limited spatial resolution and the small particle numbers.

At late times the evolution of the global star formation rate and gas surface density is qualitatively similar to that predicted by a stationary model, in which stars are formed in Toomre unstable regions on a time scale proportional to the growth time of gravitational instabilities (Wang & Silk 1994).

Since the numerical integration of the quasi–stationary state is computationally very expensive, it is important to note, that the model at $z = 1$ can be taken as being representative of the model at $z = 0$ as far as the kinematics and structure of the model are concerned. For many purposes, especially if global properties are considered, it is sufficient to perform a full simulation until $z \approx 1$ and to freeze the dynamical state afterwards. The subsequent evolution can then be simulated by only applying the star formation algorithm during the last 8 Gyr. This procedure cannot be applied, however, if the galaxy strongly interacts with the environment.

In conclusion, we are able to simulate the formation of disk galaxies, like our milky way, which allows one to get results with a numerical resolution down to kpc scales. The main properties of the computed models agree qualitatively and even almost quantitatively with those of typical observed spirals. Although the initial and boundary conditions used in the simulations are still not fully consistent with the predictions of structure formation in a hierarchical universe, we are quite optimistic that within a few years numerical simulations will become possible, which incorporate the initial conditions used by Navarro & White (1994) and at the same time have the spatial resolution of our simulations.

**Acknowledgments:** This work has been supported by a PhD scholarship of the Max–Planck–Gesellschaft and by the Deutsche Forschungsgemeinschaft under He 1487/8–1.

Figure 1:
The three panels show the radial profiles of the velocity dispersion $\sigma_r$ $(---)$, $\sigma_\phi$ $(\cdots)$ and $\sigma_z$ $(-\cdot-)$ of dark matter, young stars and old stars (from top to bottom). For comparison, the circular velocity taking into account all matter is shown in all three panels (solid line).

Figure 2:
Star formation rate in $M_\odot/yr$ as a fu.ction of look–back time.

Figure 3:
Particle distribution at the end of the simulation. On the left (right) side, particle positions are projected into (perpendicular to) the rotational plane. The upper, middle and lower panels give the distributions of the gas, the stars and the disk stars ($t_* < 8.5\,\mathrm{Gyr}$), respectively.

Figure 4:
Same as Fig. 3, but showing from top to bottom the distribution of young disk stars ($t_* < 1\,\mathrm{Gyr}$), of old metal poor (halo) stars ($t_* > 10.5\,\mathrm{Gyr}$, $Z < 0.01\,Z_\odot$) and of old metal rich (bulge) stars ($t_* > 10.5\,\mathrm{Gyr}$, $Z > 1.5\,Z_\odot$), respectively. Note the different scaling of the axes in the lower two panels.

Figure 5:
Surface density of the dark halo in a simulation without gas (left) and in one with gas and star formation (right) plotted against $(r/r_e)^{0.25}$. The effective radius $r_e$ is $35\,\mathrm{kpc}$ ($30\,\mathrm{kpc}$) for the simulation without (with) gas.

Figure 6:
Surface density for the different galactic components as a function of radius. Displayed are the surface density for three models at $z = 1.45$ (dashed), $z = 0.78$ (dotted) and $z = 0$ (solid), respectively. The different panels show the surface density for stars (upper left), gas and stars (upper right), disk stars ($t < 10\,\mathrm{Gyr}$, middle left), disk stars and gas (middle right), gas (lower left) and old stars (lower right), respectively.

Figure 7:
Mean latitude of stars as a function of their age. The full line shows the latitude distribution at the end of the simulation and the dashed line for the model at $z = 0.5$.

Figure 8:
Rotation curves of the different components of the disk galaxy. All panels show the circular velocity taking into account all matter (solid line), and only baryonic matter (dotted line), respectively. In the top panel, the rotation velocity of the gas $(---)$, of the stars $(-\cdot-)$ and of the dark matter $(-\ldots-)$ is shown in addition. The dashed dotted line in the lower two panels gives the rotation velocity of the young and old stars, respectively.